\documentstyle[prl,twocolumn,aps]{revtex}
\input{epsf}
\begin{document}
\draft
\twocolumn[\hsize\textwidth\columnwidth\hsize\csname @twocolumnfalse\endcsname
\title{Quantum Computing of Quantum Chaos in the Kicked Rotator Model}

\author{B. L\'evi, B. Georgeot and D. L. Shepelyansky}

\address {Laboratoire de Physique Quantique, UMR 5626 du CNRS, 
Universit\'e Paul Sabatier, F-31062 Toulouse Cedex 4, France}

\date{October 21, 2002}

\maketitle

\begin{abstract}
We investigate a quantum algorithm which simulates efficiently
the quantum kicked rotator model,
a system which displays rich physical properties, and enables to study
problems of quantum chaos, atomic physics and 
localization of electrons  in solids.
The effects of errors in gate operations are tested on
this algorithm in numerical simulations with up to 20 qubits.
In this way various physical quantities are investigated.  Some of them,
such as second moment of probability distribution and 
tunneling transitions through invariant curves
are shown to be particularly sensitive to errors.  However,
investigations of the fidelity and  Wigner and Husimi distributions
show that these physical quantities are robust in presence of imperfections.
This implies that the algorithm can 
simulate the dynamics of quantum chaos in presence of a
moderate amount of noise.  
\end{abstract}
\vskip1pc]

\narrowtext

\section{Introduction}

It is only recently that it was realized that quantum mechanics 
can be used to process information in fundamentally new ways.
In particular, Feynman \cite{feynman} emphasized that the massive
parallelism due to the superposition principle may allow to simulate
efficiently some problems intractable on classical
computers, the most obvious being many-body quantum systems.
Since that time, a model of quantum computer has been set up,
viewed as an ensemble of {\it n  qubits}, i.e. two-level systems,
with a Hilbert space of dimension $2^n$ (see reviews
\cite{josza,steane,nielsen}).
Computation is performed through unitary transformations 
applied to the quantum wave functions of this many-body system.
  In fact,
it has been shown that any unitary transformation on this 
$2^n$ dimensional space can be written in terms of a set
of universal gates, for example one- and two-qubit transformations.
Also, important quantum algorithms have been developed, such as Shor's
algorithm for factoring large numbers \cite{shor} which is
exponentially faster than any known classical method, and Grover's
algorithm to search a database \cite{grover}, 
where the gain is polynomial.

Motivated by these developments, many experimental implementations
for actual realization of such a quantum computer were proposed
(see \cite{nielsen} and references therein).
Recent results include for example the NMR implementation of
factorization algorithm with seven
 qubits made from nuclear spins in a molecule 
\cite{7q}, and the simulation of the quantum baker map \cite{baker}.
Thus small quantum computers with a few qubits are already 
available experimentally, and
systems of larger size can be envisioned at relatively short
term.

Still, algorithms such as the one of Shor require
large number of qubits and the use of many gates.
It is therefore important to develop algorithms which need a smaller
number of qubits and gates and still can yield interesting quantities.
In particular, algorithms enabling to simulate quantum mechanical
systems, as originally envisioned by Feynman, can be implemented 
relatively easily and solve problems inaccessible to classical computers
with less expanse in number of qubits and gates. 
Several such algorithms have been developed for various systems, including
many-body Hamiltonians \cite{lloyd} or spin lattices \cite{molmer}.
An especially interesting class of systems corresponds to
chaotic quantum maps.  Such systems can have a very complex dynamics while
their Hamiltonians keep a relatively simple form.  Algorithms for fast
simulation on a quantum computer were built for the quantum baker map
\cite{schack}, the kicked rotator \cite{GS}, and the sawtooth map
\cite{benenti}. We note that recently the quantum baker map 
was implemented on a NMR quantum computer \cite{baker}.
 The kicked rotator is an especially rich and generic system, 
which has been a cornerstone for the study of quantum chaos \cite{qchaos}.
In the classical limit it reduces to the Chirikov standard map which has been 
also extensively studied in the field of classical chaos \cite{lichtenberg}.
Implementation of this model can be done on a small quantum computer with a
few tens of qubits, and classical supercomputers will be outperformed 
with a few hundreds of qubits.  
Still, real quantum computers will not be free of imperfections and errors,
and this will affect the results of the computation. 
It is therefore important to understand the effects of different 
sources of errors on the results of such an algorithm.
 For example, first numerical
simulations of the quantum computation of this model \cite{song} have
shown that errors affect in a different way the various physical quantities
characterizing the model, and that for some of them the effect of errors can
be exponentially strong.

In this paper, after presenting in more detail the physics of the kicked 
rotator, we study the effects of errors on several physical quantities.
We focus on random unitary errors, which may arise when imperfect gates are 
applied, and study first how global quantities
such as second moment or fidelity are affected by errors.  
Our results confirm and extend those obtained
in \cite{song} showing a marked contrast in the behavior of these two
quantities in presence of errors.  We also investigate how well the whole
wave function is  reproduced by an imperfect quantum computer.
 A particularly interesting way to display wave functions
is to express them through
phase space distributions, such as Wigner and Husimi functions.  
These distributions display the same information as
the wave functions, but in a form which allows direct comparisons 
between classical and quantum dynamics, a property
especially interesting to probe the classical limit of quantum mechanics.
  They have been extensively
 used in many
fields, and recently a method has been devised \cite{paz} to measure 
such distribution for quantum simulations on quantum computers.
The effects of errors on Wigner and Husimi functions will be investigated
in details, showing how imperfections affect the different 
parts of phase space, and discussing how
information can be retrieved through quantum measurement.
 A separate section is focused on how a
localized distribution may escape from an island of integrability, 
showing an especially large effect of quantum errors on a quantity which
is directly relevant to quantum tunneling.

\section{The kicked rotator}

The classical kicked rotator is described by the Chirikov standard map
\cite{qchaos,lichtenberg}:

\begin{equation}
\label{map}
\bar{n} = n + k \sin{  \theta }; \;\;\; 
\bar{\theta} = \theta + T \bar{n}
\end{equation}

\noindent where $(n,\theta)$ is the pair of conjugated momentum (action) 
and angle 
variables, and the bars denote the resulting variables after one iteration 
of the map.  It describes a free angle rotation and a kick in momentum.
This area-preserving map has been extensively studied during
the past decades, and has been applied
to problems
such as particle confinement in magnetic traps, beam dynamics in
accelerators, comet trajectories and many others \cite{lichtenberg}.  

The dynamics of this map takes place 
on a cylinder (periodicity in $\theta$), and is controlled by
a single parameter $K=kT$. For $K=0$ the system is integrable 
and all trajectories lie on one-dimensional tori (lines $n=$constant).
For $K>0$, the system undergoes a transition to chaos, which follows
the Kolmogorov-Arnold-Moser (KAM) theorem. Periodic orbits
corresponding
to  rational frequencies are
transformed into chains of integrable islands mixed with chaotic region.
On the contrary, tori with irrational frequencies are deformed but survive,
forming invariant curves which separate zones in phase space.  As $K$ is
increased, these surviving tori become Cantor sets (cantori) and disappear.
The most robust torus corresponds to the golden number $(1+\sqrt{5})/2$,
and disappears for $K=K_g \approx 0.9716...$. Thus, for $K>K_g$ global chaos 
sets in, with appearance of an extended chaotic region in phase space,
and with dynamics characterized by a
positive Kolmogorov-Sinai entropy $h \approx \ln (K/2)>0$ (for $K\geq 6$).
In this r\'egime, a typical trajectory shows diffusive growth of momentum which
statistically can be described by the Fokker-Planck equation, with
diffusion rate $D=n^2/t
\approx k^2/2 $ where $t$ is measured in number of iterations (kicks)
\cite{qchaos,lichtenberg}.
For lower values of $K$, the phase space displays a complex hierarchical
structure with integrable islands surrounded by chaotic zones at smaller and
smaller scales. 

The map (\ref{map}) is periodic in $n$ with period $2\pi/T$, so the 
phase space structures repeat themselves on each cell of size $2\pi/T$.
Such a cell is shown on Fig.1 for $K=K_g$, displaying
the complex hierarchical structures which appear in the phase space.
\begin{figure}
\epsfxsize=2.8in
\epsfysize=2.8in
\epsffile{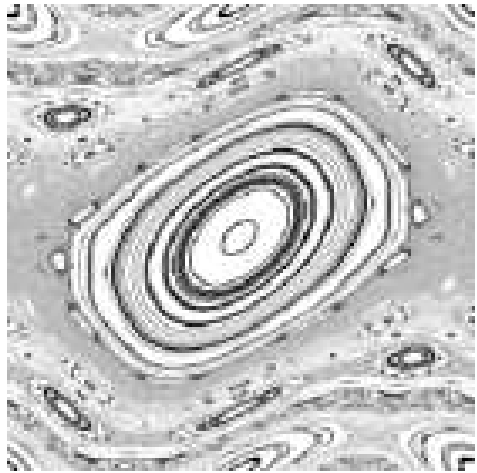}
\caption{Plot of the classical phase space at $K= \\ K_g=0.9716...$
($t=10^4$ iterations of (\ref{map}) for $200$ points). 
} 
\label{fig1}
\end{figure}

The quantization of (\ref{map}) yields a Hamiltonian which after
integration over one period gives a
unitary evolution operator
acting on the wave function $\psi$
\begin{eqnarray} 
\label{qmap}
\bar{\psi} = \hat{U} \psi =  e^{-ik\cos{\hat{\theta}}}
 e^{-iT\hat{n}^2/2} \psi,
\end{eqnarray}
where $\hat{n}=-i \partial / \partial \theta $, $\hbar=1$,
and $\psi(\theta+2\pi)=\psi(\theta)$.
The quantum dynamics depends on two parameters $k$ and $T$
(instead of the single 
parameter $K=kT$ for the classical one). 
The 
classical limit corresponds to $k \rightarrow \infty$, $T \rightarrow 0$ while
keeping $K=kT=$constant \cite{qchaos,kicked,kicked1}.  
In a sense, $T$ plays the role of an effective $\hbar$.

Depending on the values of theses
parameters, the system follows different r\'egimes, from regularity
to quantum chaos.  Due to this variety of behaviors, 
the quantum kicked rotator has been intensively studied 
(see \cite{qchaos,kicked,kicked1} and Refs. therein). Indeed, 
most of the  phenomena characteristic
of quantum chaos are present, such as quantum ergodicity, Random
Matrix Theory statistics, chaos assisted tunneling, and others.
In particular, for $K > K_g$, the phenomenon of
dynamical localization appears. Although in this r\'egime a typical classical
trajectory diffuses in momentum, the eigenstates $\chi_m(n)$ of 
the operator $\hat{U}$ in momentum space are exponentially localized
for typical values of $k$ and $T$.  
Their envelopes obey the law  $\chi_m(n) \sim \exp(-|n-m|/l)/\sqrt{l}$ 
where $m$ marks 
the center of the eigenstate and $l$ is the localization length.
For $k \gg K \gg 1$ this length is determined by the 
classical diffusion rate $l = D/2 \approx k^2/4$ \cite{kicked}.  This
phenomenon has close relationship with the Anderson localization of electrons
in disordered solids \cite{fishman}, and investigation of
 the kicked rotator
gives information on this important
solid-state problem still under intensive 
investigation nowadays.  
The quantum kicked rotator describes also the properties of microwave
ionization of Rydberg atoms \cite{IEEE}.  It
has been realized experimentally with cold atoms, 
and the effects of dynamical localization, external noise and decoherence
have been studied experimentally \cite{raizen}.

For numerical studies of the quantum evolution (\ref{qmap}) it is convenient 
to choose the case of quantum resonance with $T/(4\pi)= M/N$ where $M, N$
are integers \cite{kicked1}. In this way the quantum dynamics
takes place on a torus with $N$ levels.
For $l \gg N$ the eigenstates of evolution operator become ergodic and 
the level spacing statistics is described by random matrix theory 
\cite{kicked1}.  

The algorithm for the quantum simulation of the kicked rotator
was presented in \cite{GS}.  The evolution (\ref{qmap})
consists of the product of two unitary operators, which are diagonal
in the angle and momentum bases respectively.
The most efficient classical algorithm available consists in
changing back and forth between the angle and momentum representation
by Fast Fourier Transforms (FFT).  The operator which is diagonal in the
basis is then implemented by direct multiplication of the coefficients of the
wave function. In total, one iteration of (\ref{qmap}) on
a Hilbert space of dimension $N=2^{n_q}$ requires $O(N \mbox{log}N)$ classical
operations, the limiting steps being the FFT.
The quantum algorithm follows the classical one, and speeds
up all parts of it to obtain exponential increase of computation rate.
First an initial distribution is built, in a polynomial number of 
operations (in $n_q$).  Various initial wave functions can be built
in such a way.  In the following, we will use as initial state $|\Psi_0\rangle $ a
wave function localized at a precise value of momentum $n$, which can be built
in $n_q$ single-qubit rotations starting from the ground state. 
The general state of the system can be written as $\sum_{n=0}^{N-1} a_n |n\rangle $,
where $a_n$ are the amplitudes of the wave function 
on the $|n\rangle $ basis state.
Then the first unitary operator is applied.  
In the $n$ representation it is diagonal and can be written
$\exp(-iTn^2/2)$.  This operator can be implemented efficiently
by using the binary decomposition of $n$: if
$n=\sum_{j=0}^{n_q-1} \alpha_j 2^j$, then
$n^2= \sum_{j_1,j_2} \alpha_{j_1} \alpha_{j_2} 2^{j_1+j_2}$.  Therefore
$\exp(-iTn^2/2)=\Pi_{j_1,j_2} \exp(-iT\alpha_{j_1} \alpha_{j_2} 
2^{j_1+j_2-1})$ 
with $\alpha_{j_{1,2}}=0$ or $1$. Thus one needs to implement 
the two-qubit gate applied to each qubit pair 
$(j_1,j_2)$ which keeps the states $|00\rangle , |01\rangle , |10\rangle $ 
unchanged while $|11\rangle $ is transformed to $\exp(-iT 2^{j_1+j_2-1}) |11\rangle $.
$O(n_q^2)$ applications of this gate are sufficient
 to simulate $\exp(-iTn^2/2)$.

Then a quantum Fourier transform (QFT) (see e.g. \cite{josza}) 
is performed to shift from
$n$ to $\theta$ representation, yielding $\sum_{i=0}^{N-1} b_i |\theta_i\rangle $.
This transformation needs only
$O(n_q^2)$ one and two-qubit gates, and yields the wave function
in $\theta$ representation. In this representation, the second operator 
$\exp(-ik\cos \hat{\theta})$ is diagonal.  
Direct (sequential) multiplication by
$\exp(-ik\cos\theta_i)$ for each $\theta_i$ will require exponentially many
operations, so a parallel way to apply this operator has to be devised.
In \cite{GS}, it was proposed to use supplementary registers on which
the values of $\cos( \theta_i)$ will be computed in parallel.
The procedure transforms $\sum_{i=0}^{N-1} b_i |\theta_i\rangle |0\rangle $ into
$\sum_{i=0}^{N-1} b_i |\theta_i\rangle |\cos \theta_i\rangle $ with $\cos( \theta_i)$
computed up to a fixed precision using a recursive method based
on Moivre's formula \cite{GS}. This is actually the slowest step of
the algorithm, requiring $O(n_q^3)$ elementary operations.
From the state $\sum_{i=0}^{N-1} b_i |\theta_i\rangle |\cos \theta_i\rangle $,
it is easy by using $n_q$ one-qubit operations to build the state
$\sum_{i=0}^{N-1} b_i \exp(-ik\cos\theta_i)|\theta_i\rangle |\cos \theta_i\rangle $.
Then the cosines in the last register are reversibly erased by running
backward the sequence of gates that constructed them, and one ends up
with the state $\sum_{i=0}^{N-1} b_i \exp(-ik\cos\theta_i)|\theta_i\rangle |0\rangle $,
which is the result of
 the action of the unitary operator $\exp(-ik\cos\hat{\theta})$.
Another QFT (requiring $O(n_q^2)$ operations) takes the wave function back
to the $n$ representation.  

Increasing $n_q$, which exponentially increases the dimension of the
Hilbert space available, enables to probe various physical limits
in the system. If $K=kT$ is kept constant, the classical
mechanics remains the same.  If $T$ is kept constant,
the effective
$\hbar$ is fixed, and increasing $n_q$ will increase exponentially the size 
of the phase space of the system (number of cells).
In contrast, if $T=2\pi/N$, with $N=2^{n_q}$,
the size of the phase space remains the same, all $N$ momentum
states corresponding to the same cell of size $2\pi/T$.
In this case, increasing $n_q$ increases
the number of quantum levels corresponding to the same classical structure,
and is equivalent to decreasing $\hbar$ toward the classical limit.

The whole quantum algorithm described above
requires $O(n_q^3)$ gate operations to perform one
iteration of the quantum map (\ref{qmap}), exponentially less than the
classical algorithm.  Still, a physical quantum computer will not be
an ideal perfect machine, and there will be imperfections, which may hamper
the computation. In the following sections, we will investigate the effects
of noise and imperfections on the physical quantities that are simulated,
and estimate the accuracy of the quantum computation of the quantum 
map (\ref{qmap}). The numerical simulation
of many qubits is very resource-consuming on a classical computer.
Due to that we took in all numerical computations
the action of $\exp(-ik\cos \hat{\theta})$ as exact, and performed
by direct multiplication in $\theta$ representation, all other
operations being made with errors.  We think that
this approximation does not alter the qualitative features of the results,
although the number of quantum 
gates is reduced from $O(n_q^3)$ to $O(n_q^2)$. Also in this approximation
all supplementary registers required for the computation
of $|\cos \theta_i\rangle $ are eliminated and the quantum evolution on $N=2^{n_q}$
levels is performed with only $n_q$ qubits.

\section{Global quantities}

We first study the effects of imperfections and errors for the global
quantities of the system.
\begin{figure}
\epsfxsize=3.2in
\epsffile{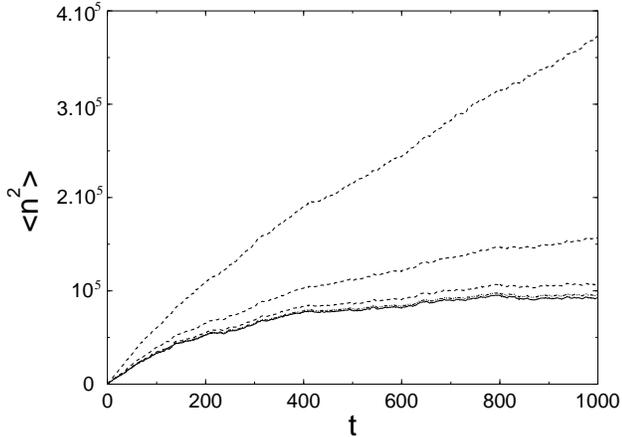}
\vglue 0.2cm
\caption{Dependence of second moment $\langle n^2\rangle =\langle (n-n_0)^2\rangle $ of the 
probability distribution on time $t$ for $T=0.5$ and $K=15$.
Data are shown from top to bottom for $n_q=16,15,14,13$  
and $\varepsilon=10^{-4}$ (four curves).
The lowest fifth full curve  is for $\varepsilon=0$, $n_q=14$.
Initial state is $|\Psi_0\rangle = |n_0\rangle $, with $n_0=N/2$.} 
\label{fig2}
\end{figure}
To model these imperfections, we introduce a random unitary error
during the operation of elementary quantum gates.  These errors are present 
for each gate performing the quantum Fourier transform and the action
of the unitary operator $e^{-iT\hat{n}^2/2}$.  Two elementary gates are used:
single-qubit Hadamard gates $H= \mbox{diag}(1,1,1,-1)$ and the two-qubit gate
$B=\mbox{diag}(1,1,1,\exp (i\alpha))$ where $\alpha$ is a phase.
The transformation $H$ can be written $H= \vec{u_0} \vec{\sigma}$
where $\vec{u_0}=(1/\sqrt{2},0,1/\sqrt{2})$ and $\vec{\sigma}=
(\sigma_x, \sigma_y, \sigma_z)$.  It is replaced by an imperfect gate
$H'=\vec{u} \vec{\sigma}$, where $\vec{u}$ is a unit vector
with a random angle $\beta$ from $\vec{u_0}$.  In a similar way,
each $B$ is replaced by $B'=\mbox{diag}(1,1,1,\exp (i\alpha+i\gamma))$ where
$\gamma$ is again a random angle.  At a given strength $\varepsilon >0 $ 
of noise, each gate is implemented with a
$\beta$ or $\gamma$ randomly selected from a
uniform distribution such that
$|\beta| < \pi \varepsilon$ or $|\gamma|< \pi \varepsilon$
\cite{notesong}.  As explained in Section II,
we made the approximation of taking the
action of $\exp(-ik\cos \hat{\theta})$ as exact, all other
operations being made with errors. The use of $n_q$ qubits
gives a Hilbert space for wave functions of the kicked rotator
on $N$ levels, with $N=2^{n_q}$, i. e. values of momentum
range from $n=1$ to $n=N$.
In all numerical computations, the 
initial state $|\Psi_0\rangle $ was chosen as localized on a precise value
of the momentum $n_0$, i. e. $|\Psi_0\rangle = |n_0\rangle $, with $n_0=1$ 
(lowest value of momentum) or $n_0=N/2$. The rotation is computed as
$\exp(-iT(n-\bar{n})^2/2)$ with $\bar{n}=N/2$.

Depending
on the choice of parameters in (\ref{qmap}), increasing the number of
qubits $n_q$ will increase the number of values of momentum in each
phase space cell of size $\Delta n =2\pi/T$, or increase the number of cells,
or both.  
In \cite{song}, it was shown that if $T$ is constant while $n_q$ increase,
errors in the QFT may lead to
an {\em exponential} growth of errors with $n_q$ for the second moment $\langle n^2 \rangle$
of the probability distribution.  In this case, the size of phase space
grows exponentially with $n_q$ but $K$ and the effective $\hbar$ are
kept fixed.  Due to quantum localization, exact wave functions cannot
spread beyond a region of size given by the localization length, which remains
fixed when $n_q$ increases.  Therefore for all values of $n_q$, the second
moment of a distribution  initially located at $n_0=N/2$ 
will saturate with time
at a value independent of $n_q$ (full line in Fig.2) if (\ref{qmap})
is exactly simulated.  On the contrary,
errors in the QFT lead to small transfer
of probabilities to regions of phase space which are exponentially
far away from where the exact wave function is localized.  This 
induces the exponential increase of
the second moment with $n_q$.
We confirm here this effect in Fig.2 for different parameters
with the more complete set of errors used in
this paper, and with simulations up to larger number of qubits.

\begin{figure}
\epsfxsize=3.2in
\epsffile{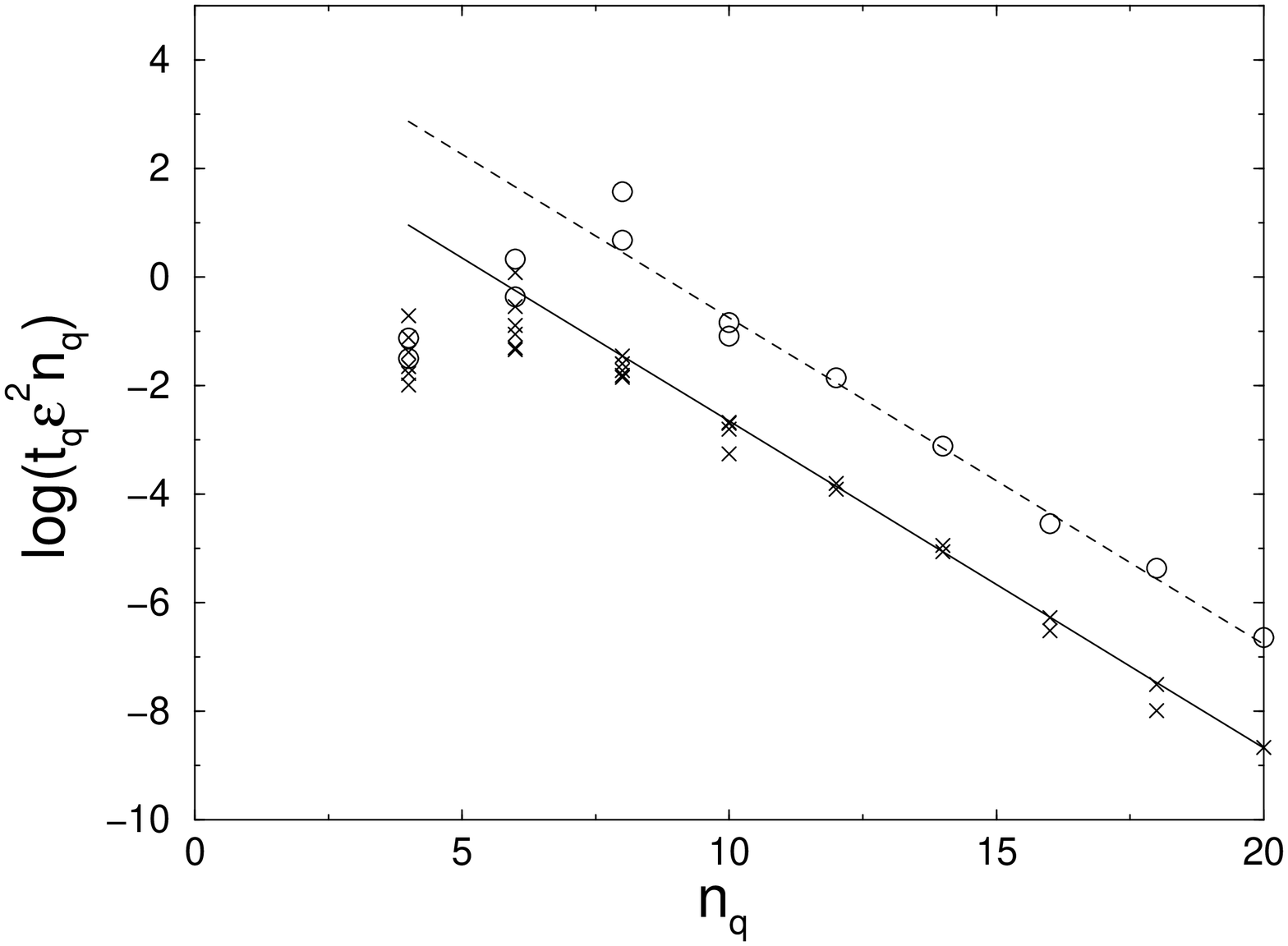}
\vglue 0.2cm
\epsfxsize=3.4in
\caption{Dependence of rescaled time scale $t_q$ on number of qubits $n_q$
for $ 10^{-6}< \varepsilon< 0.03$, $T=0.5$, $K=5$ ($\times$)
and $K=15$ ($\circ$). Initial state 
is $|\Psi_0\rangle = |n_0\rangle $ with $n_0=N/2$. Data are
averaged over $10$ to $1000$ realizations of noise.
Full and dashed lines correspond to the theoretical formula (\ref{texp})
with $C_q = 0.23$.
Logarithm is decimal.} 
\label{fig3}
\end{figure}

To be more quantitative,
Fig.3 shows the time  scale $t_q$ on which the presence of errors 
leads to a doubling of the value of the second moment $\langle n^2\rangle $ as a function
of  $n_q$ and error strength $\varepsilon$.  
In \cite{song} the formula:

\begin{equation}
\label{texp}
t_q \approx C_q k^4/(\varepsilon^2 n_q 2^{2n_q})
\end{equation}
was proposed and checked numerically with up to 13 qubits.  It stems from
the fact that each imperfect gate operation transfers on average a probability
of $\varepsilon^2$ equally divided among $n_q$ 
spurious peaks located at integer powers of $2$.
Thus due to imperfections $\langle n^2\rangle \sim n_q \varepsilon^2 2^{2n_q}t$ (each time step
involves $\sim n_q^2$ gate operations), whereas for the exact wave function
$\langle n^2\rangle \approx D^2 \approx 4l^2\approx k^4/4$. Both expressions become
comparable at the time $t_q$ given by  (\ref{texp}).
Fig.3
confirms this formula by extensive numerical computations,
with up to 20 qubits, and for two different values of $K$.
This enables to get the numerical constant $C_q\approx 0.23$.

\begin{figure}
\epsfxsize=3.2in
\epsffile{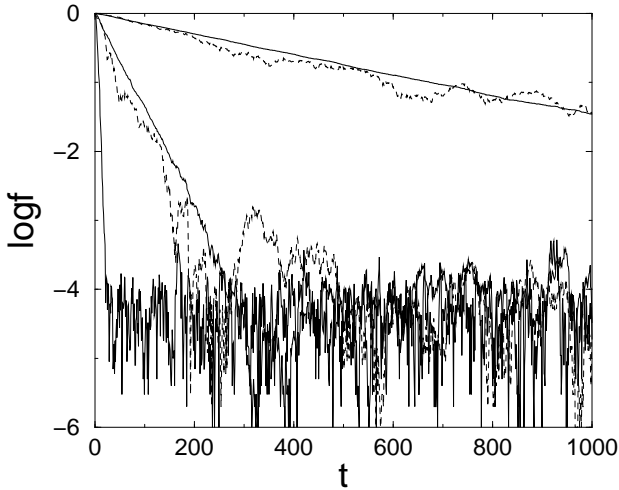}
\vglue 0.2cm
\caption{Evolution of fidelity $f$ with time $t$.
Full curves are for $K=1.3$, $T=2\pi/N$ ($N=2^{n_q}$), $n_q=14$, and from top
to bottom  $\varepsilon=3\times 10^{-3}$,
$\varepsilon=10^{-2}$, $\varepsilon=0.03$. Initial state 
is $|\Psi_0\rangle = |n_0\rangle $ with $n_0=1$.
Dashed curves are for $T=0.5$, $K=5$ and $n_q=14$, and from top
to bottom  $\varepsilon=3 \times 10^{-3}$, $\varepsilon=10^{-2}$.
Initial state is $|\Psi_0\rangle = |n_0\rangle $ with $n_0=N/2$. Logarithm is decimal.} 
\label{fig4}
\end{figure}

Although the time scale $t_q$ drops exponentially with $n_q$, there are other 
observables which show only polynomial sensitivity to errors.
A standard quantity used to characterize 
the global influence of errors is the fidelity
defined by the projection of the wave function with errors $\psi_{\varepsilon}(t)$
  on the perfect one $\psi_0 (t)$: $f(t)=|\langle \psi_{\varepsilon}(t)|\psi_0 (t)\rangle |^2$.
The dependence of this fidelity on time in presence of errors
is shown in Fig.4, showing that it decreases slowly with $t$
 and amplitude of noise $\varepsilon$.
One can define a time scale $t_f$ such that $f(t_f)=0.5$.  Fig.5
presents the variation of $t_f$ with system parameters in two
different r\'egimes. It  shows that the relation:
\begin{equation}
\label{tf}
t_f \approx C_f /(\varepsilon^2 n_q^2)
\end{equation}
holds with the numerical constant $C_f\approx 0.35$.  Fig.4 and Fig.5
are consistent
with a fidelity decay $f(t)\sim \exp (-\Gamma t)$ where 
$\Gamma \sim \varepsilon^2 n_q^2$.

The relation (\ref{tf}) can be understood from the 
following physical considerations.  Each imperfect unitary gate is
rotated by a random angle of order $\varepsilon$ from
the exact one.  Therefore a probability of order $\varepsilon^2$
is transferred from the exact state at each gate operation.
Each time step of the map (\ref{qmap}) takes $O(n_q^2)$ operations,
in the approximation which we have taken where the building of the
cosines is supposed exact.  This implies that $t_f$, which is in units of 
time steps of (\ref{qmap}), should vary as $1/(\varepsilon^2 n_q^2)$.
We expect that if the full algorithm was implemented, with the
cosines computed following the procedure explained in Section II,
a time step of  (\ref{qmap}) should take $O(n_q^3)$ operations,
and accordingly $t_f$ should vary as $1/(\varepsilon^2 n_q^3)$.

\begin{figure}
\epsfxsize=3.2in
\epsffile{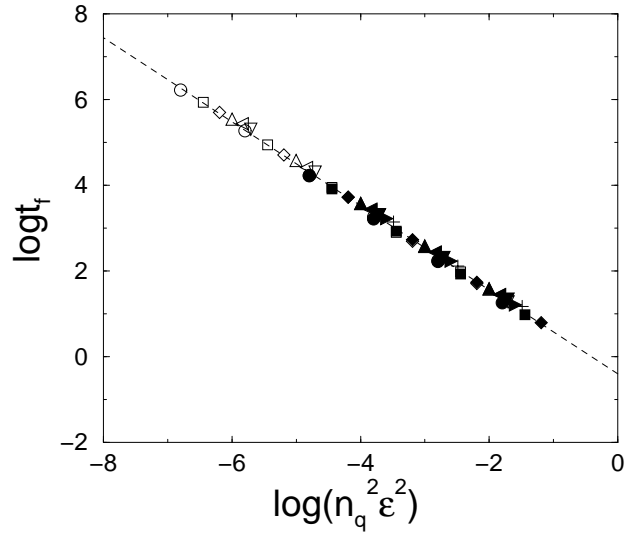}
\vglue 0.2cm
\caption{Dependence of time scale $t_f$ on system parameters
for $n_q=4$ ($\circ$), $6$ ($\Box$), $8$ ($\diamond$),$10$ ($\bigtriangleup$),
$12$ ($\triangleleft$), $14$ ($\bigtriangledown$), $16$ ($\triangleright$),
$18$ (+). Here 
$K=1.3$, $T=2\pi/N$ $(N=2^{n_q})$ (open symbols)
or $K=5$, $T=0.5$ (full symbols). Dashed line is the
theoretical formula (\ref{tf}) with $C_f=0.35$. Initial state is
$|\Psi_0\rangle = |n_0\rangle $, with $n_0=1$ ($K=1.3$) or $n_0=N/2$ ($K=5$). Data are
averaged over $10$ to $100$ realizations of noise.
Data for $T=0.5$ and $K=15$ 
are nearly indistinguishable from $T=0.5, K=5$ (not shown).
Logarithms are decimal.}
\label{fig5}
\end{figure}

The data shown in this section exemplify the sharp contrast in the
behavior of the different observables
in presence of errors.
The fidelity shows only a polynomial decrease with respect to
both $\varepsilon$ and $n_q$, whereas the second moment of
the wave function grows exponentially with $n_q$, but polynomially
with $\varepsilon$.  The resolution of this apparent paradox is related to
the fact that the second moment is sensitive to the {\em size}
of the Hilbert space, which grows exponentially with $n_q$.
Small spurious peaks due to imperfections do not spoil the fidelity,
but modify strongly the variance $\langle n^2\rangle $ if they appear 
very far away from the exact location of the wave function \cite{note}.

\section{Wigner and Husimi distributions}

In the previous section we focused mainly
on the case where $T$ (effective $\hbar$) is fixed
but the phase space size grows exponentially with $n_q$. 
In contrast, at $T=2\pi/N$
and  $N=2^{n_q}$ the system size in classical momentum 
(number of $2\pi/T$ cells in $n$) remains fixed when $n_q$ increases. 
In this way the effective $\hbar$ drops exponentially
with $n_q$ and going to larger
number of qubits means approaching the classical limit (exponentially fast).
Smaller and smaller details of the classical structure will be visible
in the quantum wave functions.  
In this r\'egime, data presented in
Fig.4 and Fig.5 have already shown that the fidelity 
follows the law (\ref{tf}) as in the case $T=$constant.
However, the fidelity characterizes in one number the
accuracy of the whole wave function, and does not tell how well the local
properties are reproduced.
To study the local properties of wave functions, 
one can express it in $\theta$ or $n$ representation.  However, a very useful
representation corresponds to phase space distributions, such as Wigner
or Husimi distributions.
They are especially used in the field of 
quantum chaos, since such representations permit a direct comparison
with classical Hamiltonian mechanics which takes place in phase space.
They also enable to probe the classical/quantum border when $\hbar$ is decreased
compare to other parameters of the system.  Plotting such quantities
in presence of errors allows to probe how local properties of the
wave functions are sensitive to imperfections in the quantum algorithm.

An additional motivation to study such phase space representations
stems from the fact that recently an algorithm was proposed  
\cite{paz} which enables to compute the Wigner function on
a chosen point in phase space by the use of an ancilla qubit.

For a continuous system with two conjugate variables $p$ and $q$
the Wigner transform \cite{wigner} of a wave function $\psi$ is defined by:
\begin{equation}
\label{wigner}
W(p,q)= \int
\frac{e^{-\frac{i}{\hbar}p.q'}}{\sqrt{2\pi\hbar}}
\psi(q+\frac{q'}{2})^{*}\psi(q-\frac{q'}{2}) dq'
\end{equation}
 
In a discrete system with $N$-dimensional Hilbert space,
 one is led to define the Wigner function on a lattice of
$2N \times 2N$ points
(see e. g. \cite{discrete}).  
In the case of the kicked rotator, the formula becomes:

\begin{equation}
\label{wigner2}
W(\theta,n)= 
\sum_{m=0}^{N-1} \frac{e^{-\frac{2i\pi}{N}n(m-\Theta/2)}}{2N}
\psi(\Theta-m)^{*}\psi(m),
\end{equation}
with $\Theta=\frac{N\theta}{2\pi}$.
The Wigner function is always real, but contrary
to classical Liouville phase space distributions it can take
negative values.  It verifies $\sum_{i} W(\theta_i,n)=|\psi(n)|^2$
and  $\sum_{i} W(\theta,n_i)=|\psi(\theta)|^2$.

\begin{figure}
\epsfxsize=3in
\epsfysize=4.5in
\epsffile{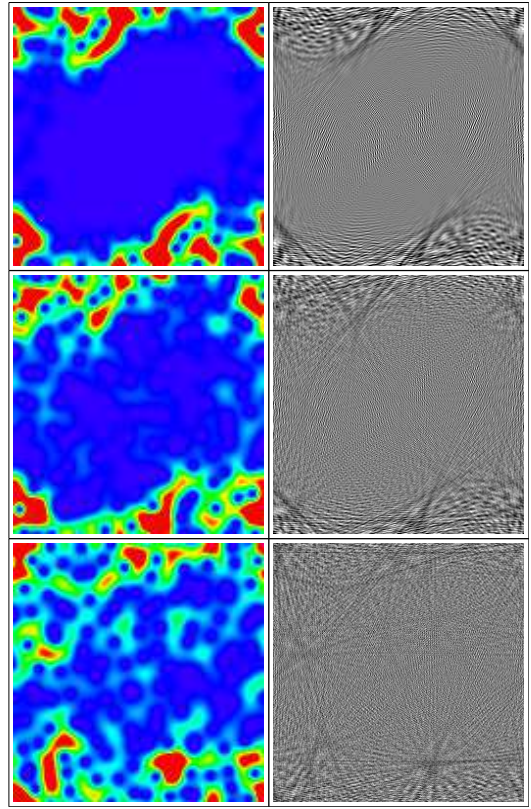}
\vglue -0.2cm
\caption{(color) Plot of Husimi (left) and
Wigner (right) distributions at $t=10^3$
for $K=1.3>K_g$, $T=2\pi/N$, $N=2^{n_q}$ and $n_q=7$.
Initial state is
$|\Psi_0\rangle = |n_0\rangle $, with $n_0=1$.
Top: $\varepsilon=0$; middle: $\varepsilon=0.002$; bottom: $\varepsilon=0.004$.
Left: color/grayness represents intensity level from blue/white (minimal) 
to red/black (maximal).
Right: grayness represents amplitude of the Wigner function,
from white (minimal negative value) to black (maximal positive value).} 
\label{fig6}
\end{figure}

The Wigner transform has the drawback of being negative or positive.
Nevertheless, coarse-graining this function over cells of
size $\hbar$ gives non negative values.  Such a procedure
gives the {\em Husimi distribution} (see e.g. \cite{husimi}) which
corresponds to a Gaussian smoothing of the Wigner function.
In the case of the kicked rotator, the Husimi distribution can be computed 
through:

\begin{equation}
\label{husimi}
h(\theta,n)= \sum_{m=n-N/2}^{n+N/2}
(\frac{T}{\pi})^{\frac{1}{4}}
\frac{\psi(m)}{\sqrt{N}} e^{-\frac{T}{2}(m-n)^2} e^{im\theta}
\end{equation}
where the Gaussian for simplicity 
is truncated for values larger than $N/2$, and $ \psi(m)$
is the wave function in momentum representation.
The Husimi distribution is always non negative, and allows
a direct comparison between classical Liouville density
distributions and 
quantum wave functions. 

Wigner and Husimi distributions of wave functions of the
quantum kicked rotator 
simulated on a quantum computer are shown on Fig.6
for different level of errors.  Both functions have similar
patterns, although as expected the Wigner function 
displays interference structures absent in the Husimi distribution.
In the regime of parameters studied, classical invariant curves
are still present in phase space and prevent the exact wave function to 
enter the large elliptical island in the middle.  In the presence of moderate
level of noise, main structures are still present and distinguishable.

\begin{figure}
\epsfxsize=3.3in
\epsfysize=4.4in
\epsffile{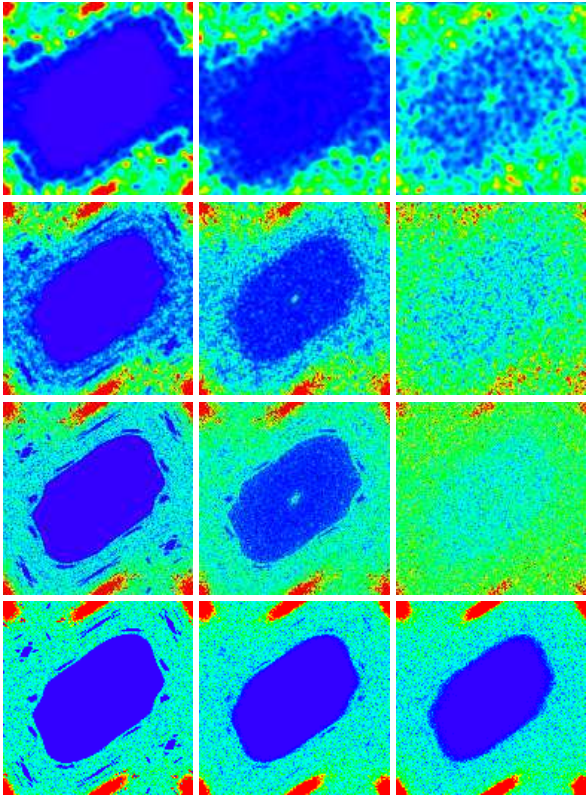}
\vglue 0.2cm
\caption{(color) First three rows:
Husimi distribution at $t=10^3$ for $K=1.3>K_g$ and 
$T=2\pi/N$, $N=2^{n_q}$; from top to bottom:
$n_q=9$, $n_q=12$, $n_q=14$; quantum noise $\varepsilon=0$ (left),
$\varepsilon=0.002$ (center), $\varepsilon=0.004$ (right). Bottom row:
classical phase space distribution at $t=10^3$
with classical noise $\varepsilon=0$ (left),
$\varepsilon=0.002$ (center), $\varepsilon=0.004$ (right).
For clarity, the distributions are averaged over $10$ iterations
around $t=10^3$. Initial quantum/classical state is $n_0=1$.  
Color/grayness represents intensity level from blue/white (minimal) 
to red/black (maximal).}
\label{fig7}
\end{figure}

Fig. 7 confirms this result, showing the Husimi distribution
for larger number of qubits, together with the classical 
phase space distribution.  The Husimi distributions in phase space show
features mimicking the classical phase space distributions,
in accordance with the correspondence principle. 
Fig.7 (left) shows that when $n_q$ is changed,
finer and finer details
of the classical structures are visible in the exact 
quantum wave function, in accordance with the fact that increasing $n_q$ 
amounts to reduce $\hbar$ and approach the classical limit.
The same figure shows that the wave function 
is spread over a larger domain of phase space as $n_q$ increases.
This can be explained by the following effect.  
In this mixed r\'egime between integrability and strong 
chaos at $K=1.3$ the invariant classical curves which
prevent any transport are no longer present since the last one
is destroyed at $K=K_g=0.97..$.  But {\em cantori} are present, 
which are remnants of the disappeared invariant curves.  They have a
fractal structure, and a wave packet can cross them only if
the holes are large enough.  These holes scale as $(K-K_g)^3$
and become comparable with the minimal
area scale of the Husimi distribution determined
by the effective $\hbar$ given by $T$. Hence, for $K-K_g \ll 1$,
the wave function is prevented to cross the cantorus for
$(K-K_g)^3 < T$.  Due to that quantum interference prevents 
the transport via cantori \cite{qchaos,kicked,geisel}.

The quantum Husimi distributions shown in Fig. 6 and Fig.7 display structures
of increasing complexity with larger $n_q$.   Still, with moderate
level of noise, the quantum computer is able to reproduce the exact
distributions with reasonable accuracy. For larger errors in
gate operations, significant probability is present
at wrong phase space locations, and phase space structures
 become blurred.
The comparison with the effect of classical noise visible in Fig.7
shows that in this case the quantum errors enable the wave function to
enter classically forbidden zones much faster, a fact which will be
analyzed in more details in the next section.  

\begin{figure}
\epsfxsize=3.2in
\epsffile{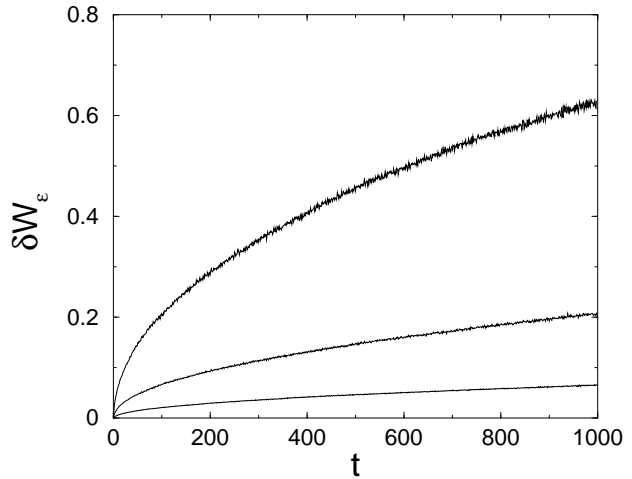}
\vglue 0.2cm
\caption{Relative error on the Wigner function $\delta W_\varepsilon=
\langle | W-W_\varepsilon| \rangle/\langle|W|\rangle $
as a function of time
for $K=K_g$, $T=2\pi/N$, $N=2^{n_q}$ and $n_q=10$.
Initial state is
$|\Psi_0\rangle = |n_0\rangle $, with $n_0=N/2$.
From bottom to top quantum noise is
 $\varepsilon=10^{-4}$, $\varepsilon=10^{-3.5}$, $\varepsilon=10^{-3}$.
The Wigner function is averaged over $2N$ values
in the chaotic zone.Data are
averaged over $10$ realizations of noise.
}
\label{fig8}
\end{figure}

It is interesting to evaluate the effects of 
noise and imperfections not only on the broad features of the full Wigner
function, but also on individual values.
In Fig.8 and Fig.9,
the behavior of individual values of the Wigner function  in presence
of noise in the gates is investigated.  Fig.8 shows that the 
{\em relative} error (i.e. the error $\langle|W -W_\varepsilon|\rangle$ 
divided by the
average individual value of the exact Wigner function $\langle |W|\rangle$)
increases slowly with the growth of $t$ and $\varepsilon$ even in
the chaotic zone.  Similar results can be observed in the integrable
zone and in the localized r\'egime (data not shown).
 In a more quantitative way, Fig.9
shows the behavior of the time scale $t_W$ when the error
on the Wigner function become comparable to its mean value
in the r\'egime chosen ($\langle|W(t_W)-W_\varepsilon (t_W)|\rangle=
\langle|W|\rangle/2$).
In all three cases considered, one obtains
\begin{equation}
\label{tW}
t_W \approx C_W/(n_q^\alpha \varepsilon^2),
\end{equation}
with $\alpha=1$ or $\alpha=1.5$.  
Thus individual values of the Wigner function in the 
kicked rotator model are robust quantities with respect to noise,
even in the chaotic r\'egime.
These results are interesting also in view of the recent discussion on the
effects of decoherence on Wigner functions \cite{zursre}.  Our results clearly 
show that in the framework of quantum computation, the errors on the
Wigner function are polynomial and not exponential.

\begin{figure}
\epsfxsize=3.2in
\epsffile{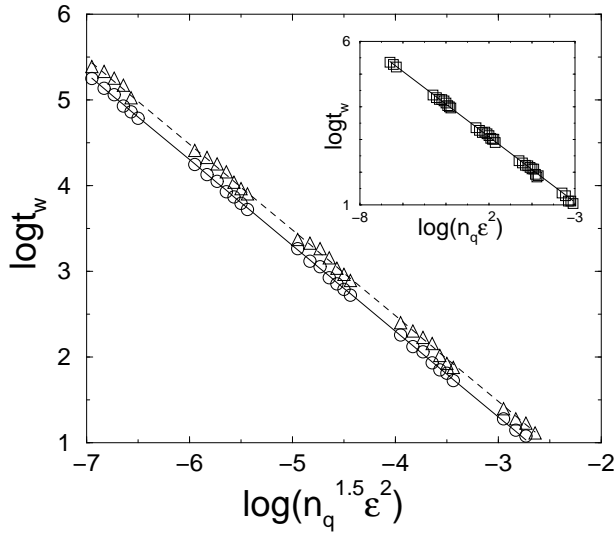}
\vglue 0.2cm
\caption{Dependence of time scale $t_W$ on system parameters
for $5\leq n_q \leq 11$. Here 
$K=K_g$, $T=2\pi/N$ $(N=2^{n_q})$. The Wigner function is averaged over 
$2N$ values
in the chaotic zone ($\bigcirc$) or in the integrable zone ($\bigtriangleup$).
Straight lines are the
theoretical formula (\ref{tW}) with $\alpha=1.5$ and $C_W=0.02$
(full line) or $C_W=0.03$ (dashed line). Initial state is
$|\Psi_0\rangle = |n_0\rangle $, with $n_0=N/2$. Data are
averaged over $10$ to $1000$ realizations of noise.
Inset: Dependence of time scale $t_W$ on system parameters
for $5\leq n_q \leq 14$. Here $T=0.5$ and $K=5$.
 The Wigner function is averaged over 
$2N$ values
in the localized zone ($\Box$).
Full line is the
theoretical formula (\ref{tW}) with $\alpha=1$ and $C_W=0.012$. Initial 
state is
$|\Psi_0\rangle = |n_0\rangle $, with $n_0=N/2$. Data are
averaged over $10$ to $1000$ realizations of noise.
Logarithms are decimal.
}
\label{fig9}
\end{figure}

As noted previously,  a recent algorithm
\cite{paz} enables to measure the value of the Wigner function 
of a system of density matrix $\rho$ on a
selected point in phase space, with the help
of an ancilla qubit $a$. 
First $H$ (Hadamard gate) is applied on $a$,
followed by a controlled-$U$ operation ($U$ is applied to the system
to be measured depending on the state of $a$) and again $H$ 
is applied on $a$.
Then the expectation value of $a$ is $\langle \sigma^z\rangle =Re[Tr(U\rho)]$.
The use of a particular operator $U$, which can be implemented efficiently
\cite{paz},
 enables to get $W(p,q)=\langle \sigma^z\rangle /2N$ (where $N=2^{n_q}$).

\begin{figure}
\epsfxsize=3.2in
\epsffile{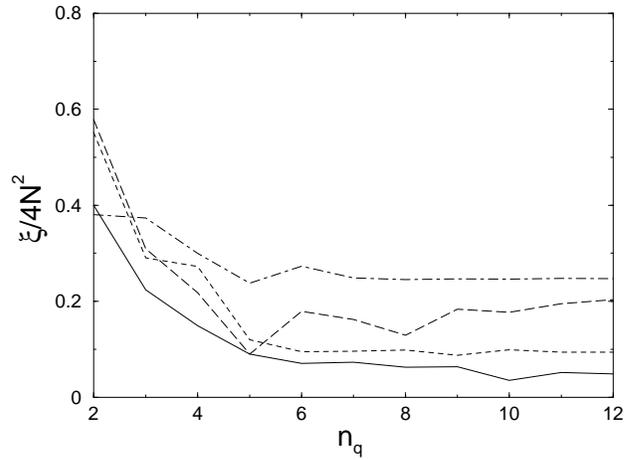}
\vglue 0.2cm
\caption{Dependence of
 inverse participation ratio $\xi$ of the Wigner
function on number of qubits $n_q$ 
at $t=10^3$ for $T=2\pi/N$, $N=2^{n_q}$ and $K=0.5$ (full curve), 
$K=0.9$ (dashed curve), $K=1.3$ (long-dashed curve), $K=2.0$ 
(dot-dashed curve). Initial state is
$|\Psi_0\rangle = |n_0\rangle $, with $n_0=1$, and $\varepsilon=0$.
} 
\label{fig10}
\end{figure}

\begin{figure}
\epsfxsize=3.2in
\epsffile{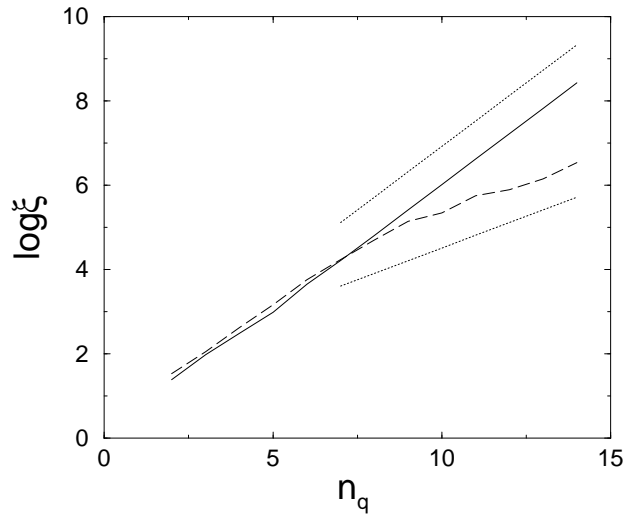}
\vglue 0.2cm
\caption{Dependence of
 inverse participation ratio $\xi$ of the Wigner
function on number of qubits $n_q$ at $t=10^3$ 
for $T=2\pi/N$, $N=2^{n_q}$ and $K=2$ 
(full line), 
and $T=0.5$ and $K=5$ (dashed line).  Dotted lines 
show $\xi\propto N^2$ and $\xi\propto N$.
Initial state is
$|\Psi_0\rangle = |n_0\rangle $, with $n_0=1$, and $\varepsilon=0$. Logarithm is decimal.} 
\label{fig11}
\end{figure}

However, we should note that even if $U$ can be 
implemented efficiently, $\langle \sigma^z\rangle $ can
be evaluated only by iterating the procedure enough times
to get a good estimate.  Therefore the amplitude of the signal
is crucial to make the whole process efficient.
Thus it is interesting to study the amplitude of peaks in the Wigner 
function, in order to know if strong peaks are present which can be 
detected reliably through this method.
This can be investigated through
a quantity which we call inverse participation ratio of the
Wigner function, in analogy with the inverse
participation ratio for wave functions used in quantum chaos
and systems with Anderson localization \cite{kramer}.  For a wave
function with $N$ projections $\psi_i$ on some basis, 
the inverse participation ratio 
$\sum |\psi_i|^2/(\sum |\psi_i|^4)$ measures the number of significant 
components in this basis.  For the Wigner function,
one has the additional sum rules $\sum W_i=1$ and $\sum W_i^2=\frac{1}{N}$.
To define an inverse participation ratio for the Wigner function,
we therefore use the formula $\xi=1/(N^2\sum W_i^4)$.
If $N$ peaks of approximately equal weights
$1/N$ are present, then $\xi=N$,
whereas $N^2$ components of equal weights (in absolute value)
$1/N^{3/2}$
give $\xi=N^2$.  This quantity $\xi$ therefore permits to estimate
the number of main components of the Wigner function. Fig.10 and Fig.11
show the scaling of this
quantity with $n_q$ for different values of parameters. 
In all the cases where $T=2\pi/N$ ($N=2^{n_q}$) the ratio
$\xi/N^2$ reaches a saturation value.
This implies that asymptotically $\langle \sigma_z\rangle =NW(p,q)\sim 1/\sqrt{N}$, 
a value
which requires $N$ iterations followed by measurements
to be reliably estimated.  In this case, the asymptotic gain in number of
operations compared
to the classical algorithm is only $O(\log (N))$, although
the resources needed are exponentially smaller ($n_q$ qubits
instead of $2^{n_q}$ classical registers).
  This should be contrasted with the case where the number
of cells increases ($T$ constant) where Fig.11 shows that 
$\xi\sim N$.  This gives $\langle \sigma_z\rangle =NW(p,q)\sim 1$, which means
that in this regime with localization, any of
the $\sim N$ components of the Wigner function which are
important can be estimated reliably and efficiently 
through this method (provided one knows beforehand
the approximate position of the localized state).
The results presented in Fig.8 and Fig.9 show that despite the different
scaling laws of $\langle |W|\rangle$, the relative errors grow
only polynomially in all cases considered, thus enabling such
measurements of individual values of $W$ to be reliable
for moderate amounts of noise.

\section{Quantum tunneling through invariant curves}

In the previous section, it was shown that the classical and
quantum errors affect the dynamics in a rather different way.  
This difference is particularly striking
in the
r\'egime where classical invariant curves are present (integrable or mixed 
systems, which correspond to moderate values of $K$ here, as in Fig.7).
Such invariant curves cannot be crossed classically, and only 
quantum tunneling can transfer probability inside 
integrable islands from chaotic regions.
However, whereas small classical errors
enable to cross only neighboring invariant curves, small quantum errors
may lead to long-distance ``jumps'' of probability deep into integrable island
(see Fig.7, last column).

\begin{figure}
\epsfxsize=3.2in
\epsffile{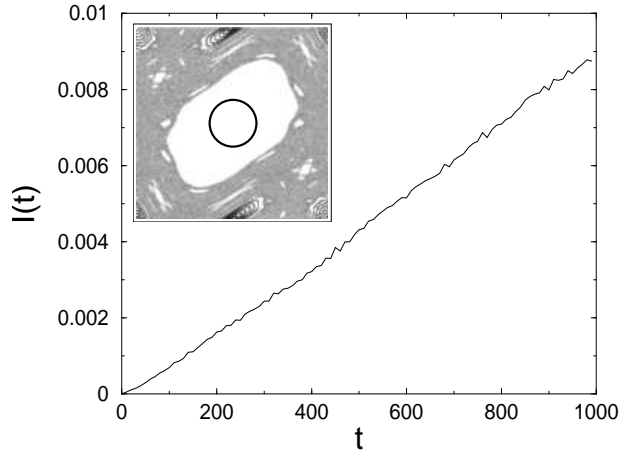}
\vglue 0.2cm
\caption{Dependence of the probability $I$ of the Husimi distribution
inside the circle (see text and inset) on time $t$
for $ \varepsilon=10^{-3}$ and $n_q=14$ at $K=1.3$ 
and $T=2\pi/N$ ($N=2^{n_q}$). Initial state is
$|\Psi_0\rangle = |n_0\rangle $, with $n_0=1$. Inset: position of
$100$ points initially at $n_0=1$ after $10^4$ iterations
of the classical map (\ref{map}), and location of the circular domain
$D$ (see text).
} 
\label{fig12}
\end{figure}

\begin{figure}
\epsfxsize=3.2in
\epsffile{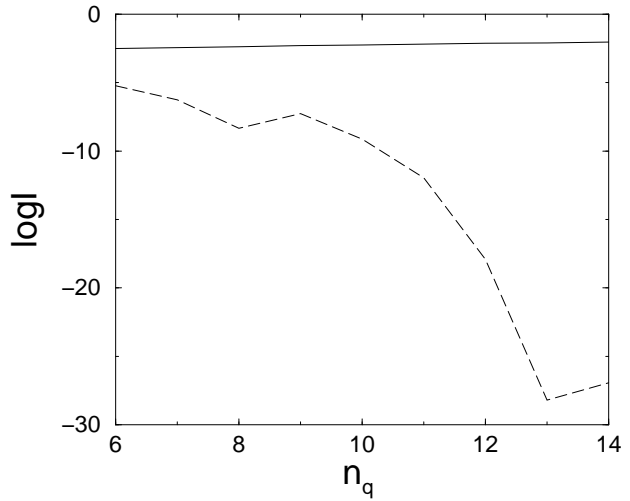}
\vglue 0.2cm
\caption{Dependence  of the probability $I$ of the Husimi distribution
inside the circle (see text and Fig.12) on $n_q$ for $K=1.3$ and
$T=2\pi/N$, $N=2^{n_q}$,
$ \varepsilon=3 \times 10^{-3}$ (solid curve) and $ \varepsilon=0$ (dashed curve). 
Data are averaged over $100$ iterations around $t=10^3$. Initial state is
$|\Psi_0\rangle = |n_0\rangle $ with $n_0=1$. Logarithm is decimal.} 
\label{fig13}
\end{figure}

To study the effects of errors on quantum tunneling we show in
Fig.12 the dependence of probability of Husimi distribution $h(\theta,n)$
inside the classically forbidden region on time $t$.
The quantity $I(t)=\int_D h(\theta,n) d\theta dn $, where $D$ is the domain 
enclosed by the circle in Fig.10 (inset), shows a linear growth with $t$.
This can be understood by a physical argument similar to the one
justifying (\ref{tf}).  Indeed, imperfect gates transfer on average
a probability of order $\varepsilon^2$ from the exact wave function to wrong phase
space positions.  However, not all gates will transfer probability
inside $D$ but only a subset of them.  This predicts that
$I(t)\sim n_q^{\alpha} \varepsilon^2 t$.  Data from Fig.12 and Fig.13 
and additional
data (not shown) confirm this prediction, with $\alpha \approx 1.3$.

To exemplify the effect of quantum errors, 
Fig.13 shows $I(t)$ at fixed time $t$
as a function of number of qubits $n_q$
for zero and nonzero noise in the gates.  In the case of 
zero noise, there is an exponential {\em decrease} with
$n_q$. Indeed, the only process which allows to enter the island for
the wave packet is quantum tunneling. In general, 
the probability of such a transition
scales like $\exp(-S/\hbar)$ where $S$ is a classical action.
 Increase of $n_q$ amounts to decrease
the effective $\hbar$  and leads to the exponential  drop
of $I$ obtained numerically at $\varepsilon=0$.  
In sharp contrast, the presence of imperfections
in the gates ($\varepsilon > 0$) leads to direct
jumps inside the island that gives an {\em increase} of $I$ with 
$n_q$ according to the estimate of the previous paragraph.
Thus for this specific process, the effect of noise in the gates results in
a qualitative change of the dependence of tunneling probability $I$
on $n_q$. 

\section{Conclusion}

The results presented in this paper show that it is possible to simulate
efficiently the quantum kicked rotator on a quantum computer.  
For the quantum algorithm simulating the dynamics of kicked rotator we
investigated the effects of gate errors and showed that certain quantities like
fidelity, Wigner and Husimi distributions are sufficiently robust
against noise in the gates. Thus for small amplitude of noise these 
quantities  can be
computed reliably without application of quantum error corrections.
At the same time we found that there exist other characteristics,
e.g. variance of probability distribution and tunneling probability inside
stability islands, which are very sensitive to errors in quantum gates.
In addition, the study of the Wigner function shows that individual
values of this function are robust with respect to 
quantum errors and can be reliably estimated. 
However, the computation of the Wigner function at specific points
meets certain readout problems in deep quasiclassical regime
where generally a large number of measurements is required.

On the basis of obtained results we believe
that the  quantum algorithms simulating quantum chaotic maps
will provide important grounds for testing the accuracy 
of the next generation of experimental implementations of 
quantum computers.

We thank the IDRIS in Orsay and CalMiP in Toulouse for 
access to their supercomputers.
This work was supported in part by the NSA
and ARDA under ARO contract No. DAAD19-01-1-0553, 
by the EC  RTN contract HPRN-CT-2000-0156 and by the 
project EDIQIP of the IST-FET programme of the EC.


\begin{thebibliography}{99}
\bibitem{notesong} In \cite{song}
               the noise amplitude was also $\pi \varepsilon$.
\bibitem{note} We note that the second moment does not grow exponentially 
            with time at fixed $n_q$, but linearly.  However, the coefficient
            of this linear growth depends exponentially on $n_q$.
\end{thebibliography}
\end{document}